\documentclass[12pt]{iopart}
\newcommand{\beq}{\begin{equation}}
\newcommand{\eeq}{\end{equation}}
\newcommand{\beqa}{\begin{eqnarray}}
\newcommand{\eeqa}{\end{eqnarray}}
\newcommand{\ba}{\begin{array}}
\newcommand{\ea}{\end{array}}
\usepackage{color}
\usepackage[dvips]{epsfig}
\usepackage{amssymb}

\usepackage{graphicx}
\usepackage{dcolumn}
\usepackage{bm}
\usepackage{epsfig}
\usepackage{amsfonts}
\usepackage{color}
\usepackage{iopams}
\usepackage{cases}

\begin{document}

\title{Localization-delocalization transition\\ of dipolar bosons in a four-well potential}

\author{G. Mazzarella$^1$, V. Penna$^2$}

\address{$^1$ Dipartimento di Fisica e Astronomia â€Galileo Galileiâ€ and
CNISM, Universit$\grave{a}$ di Padova, Via Marzolo 8, 35122 Padova, Italy\\
$^2$
Dipartimento di Scienza Applicata e Tecnologia and u.d.r. CNISM, Politecnico di Torino,
Corso Duca degli Abruzzi 24, I-10129 Torino, Italy}
\date{\today}

\begin{abstract}
We study interacting dipolar atomic bosons in a four-well potential within a ring geometry 
and outline how a four-site Bose-Hubbard (BH) model including next-nearest-neighbor 
interaction terms can be derived for the above four-well system. We analyze the ground 
state of dipolar bosons by varying the strength of the interaction between particles in
next-nearest-neighbor wells.
We perform this analysis both numerically and analytically by reformulating the dipolar-boson model within the continuous variable picture applied in [Phys. Rev. A {\bf 84}, 061601(R) (2011)]. By using this approach we obtain an effective description of the transition mechanism and show that when the next-nearest-neighbor interaction crosses a precise value of the on-site interaction, the ground state exhibits a change from the uniform state (delocalization regime) to a macroscopic two-pulse state, with strongly localized bosons (localization regime). These predictions are confirmed by the results obtained by diagonalizing numerically the four-site BH Hamiltonian.


\end{abstract}


\maketitle

\section{Introduction}
%

Dipolar quantum gases \cite{blochrev} confined in a multiple-well geometry are attracting
growing attention \cite{baranov}-\cite{maik1} due to the considerably rich scenario of novel
properties and effects that emerges from the interplay of anisotropic dipole-dipole interactions 
(coupling the magnetic/electric moments of dipolar bosons) with two-boson contact interactions 
and the interwell boson tunneling.
%

In this class of systems, special interest has been focused in the last decade on 
the simple model where bosons are trapped by a triple-well potential.
This system, effectively described by a 3-site extended Bose-Hubbard (BH) model, combines 
the effect of long-range dipolar interactions \cite{Gal}, \cite{Lu}
with the nontrivial, highly nonlinear dynamics of BH models \cite{vittorio0}-\cite{liu}.
Note that the description by means of the BH model for bosons in multi-well systems 
is reliable within certain conditions on the number of particles and the strength of the
dipolar interaction which has to not dominate the contact interaction \cite{peter,wuner}.

More specifically, in the presence of the open-chain geometry, the BH triple well has made evident the non-local character of dipolar interactions within the Josephson-like dynamics \cite{lahaye2} (in the supplemental material of the latter reference, Lahaye and co-workers have considered a four-site square system to discuss the realization of interferometric arrangements) and the possibility to induce macroscopic interwell coherence independent from the tunneling parameter \cite{fischer}.
This system has revealed as well a complex ground-state phase diagram where unstable regimes can be controlled through the dipolar and contact interactions \cite{peter}. However, it is worth to observe that this is true as long as the s-wave scattering length (which characterizes the contact interaction) is larger than a critical value depending on the geometry of the external potential and the strength of the dipolar interaction \cite{koch}.

By adopting instead the ring geometry (closed chain with periodic boundary conditions) one finds that,
in addition to translational invariant vortex-like states \cite{lasph}, \cite{paraoanu}
arising when the system includes only contact interactions, the presence of dipolar
interactions shows the formation of different density-wave states
\cite{maik1} and the possibility to observe the transition between them.
Recently, the ground-state phase diagram of the closed BH triple well has been explored
to show the influence of the possible anisotropy of dipolar interactions \cite{Gal}
while the coherent control of boson tunneling through dipolar interactions has been
studied in the presence of high-frequency time-periodic local potentials \cite{Lu}.
%
In the recent paper \cite{dmps} the ground-state properties of dipolar bosons
trapped in a 3-well potential have been investigated when the on-site interaction $U_0$
and the dipolar interaction $U_1$ are varied. The ring geometry assumed for this model
has been used to show its complete equivalence with the symmetric 3-site BH model.
The nice result was that the term representing dipolar interactions can be absorbed
in the on-site interaction term of the equivalent BH model whose strength has the form
$U= U_0-U_1$. This equivalence has allowed one to exploit the considerable amount of
information about the low-energy properties of BH model to investigate the 3-well dipolar model.
In particular, by varying $U_0$ and $U_1$, the ground state of dipolar bosons
has been found to involve dramatic changes of their space distribution
which, within the equivalent BH picture, are caused by the change of the on-site interaction
from attractive ($U<0$) to repulsive ($U>0$).
The corresponding entanglement properties have been explored.

In this work, we consider interacting dipolar bosons at zero temperature confined in a 4-well
potential forming an equilateral square. The microscopic dynamics of this system is still
described by a 4-site extended Bose-Hubbard (EBH) Hamiltonian which includes the
hopping processes through the amplitude $J$ and the same on-site effective interaction
$U= U_0-U_1$ used for the 3-well dipolar model. The novel aspect
is that the apparently harmless addition of the fourth well totally changes the symmetry
properties of the dipolar model. This causes the occurrence in the equivalent BH model of
an extra-term in which non adjacent sites feature a dipolar-like interaction term depending on
$U_1$. Such a term does not occur in the BH model related to the 3-well dipolar model. Its
presence dramatically changes the properties of the ground state.

To explore the new scenario we resort to the semiquantum approach applied in several
papers (see, e.g., Refs. \cite{Java}-\cite{vittorio2}) which reduces the Schr\"odinger 
problem of many-boson models to a diagonalizable form. A ``dual" version of this method
is also known which has been developed for spin models and applied to two-mode bosonic 
systems \cite{jpb46}.
In the sequel, we refer to this method as the continuous variable picture (CVP).
%
The latter allows one to derive a suitable set of equations describing the boson populations 
of low-energy states and to exploit their solution to reconstruct the low-energy eigenstates 
of the system. This diagonalization scheme has been successfully applied to highlight the inner
mechanism governing the localization-delocalization transition characterizing the BH models with
attractive interaction \cite{vittorio2}, \cite{vittorio1}. 

Within the CVP framework the ground
state structure is predicted to exhibit a change when $U_1$ becomes larger than $U_0/2$. 
The regime characterized by the uniform-boson distribution (delocalization regime) transforms into
a non uniform distribution (localization regime) where the ground state features the almost complete
boson-localization in two non adjacent wells. The so predicted delocalization-localization 
transition is corroborated by the results deriving from the numerical diagonalization of the 
four-site BH Hamiltonian. Thanks to the numerical approach one observes that the delocalization 
regime corresponds to a Fock state with the bosons equally shared among the four wells. 
The localization regime corresponds, instead, to a symmetric superposition of two 
Fock states each one characterized by non adjacent wells occupied by half of the total 
boson-population.
It is interesting to observe that the emergence of the two-pulse state as ground state 
allows us to establish an immediate link with the mechanism responsible for the occurrence 
of the checkerboard-insulator in optical lattices (that, actually, can be regarded as 
multi-well systems) \cite{goral}.
%

\section{The model Hamiltonian}

The model describing $N$ dipolar interacting bosons trapped by a potential $V_{t}({\bf r})$
can be derived from the bosonic-field Hamiltonian
\beqa
&&
\hat{H} =\int d^{3} {\bf r}\hat{\Psi}^{\dagger}({\bf r})\,H_0\,\hat{\Psi}({\bf r})
\nonumber\\
&+&\frac{1}{2}\int d^{3}{\bf r}\,d^{3} {\bf r'}\hat{\Psi}^{\dagger}({\bf r})
\hat{\Psi}^{\dagger}({\bf r'})V({\bf r}-{\bf r'})\hat{\Psi}({\bf r'})\hat{\Psi}({\bf r})
\label{secondq}
\, ,
\eeqa
where $\hat{\Psi}({\bf r})$ is the bosonic field,
$H_{0}={\bf p}^2/(2m) +V_{t}({\bf r})$, ${\bf p}= -i\hbar \nabla$, and
$m$ is the boson mass.
The trapping potential
\beqa
\label{quadruple}
V_{t}({\bf r})= \frac{m}{2}\,\omega_{z}^2 z^2
-V_{0}\, \sum^L_{i=1} \exp\bigg(-\frac{2\,({\vec r} - {\vec r}_i )^2 }{w^2}\bigg)\, ,
\eeqa
characterized by the trapping frequency $\omega_z$ in the axial direction, represents
the superposition of a strong harmonic confinement along axis $z$ with $L$ (planar)
potential wells placed at the equidistant sites ${\vec r}_i$ of a ring lattice.
In the presence of four potential wells the lattice is a square with side
$\sqrt 2 \ell$ and vertices ${\vec r}_1 = (\ell, 0)= - {\vec r}_3$ and
${\vec r}_2 = (0, \ell)= -{\vec r}_4$.  $V_0$ is the depth of each well.
The bosonic field $\hat{\Psi}({\bf r})$ can be expanded in terms of the annihilation operators $\hat{a}_{k}$
\beq
\label{bosonicfield}
\hat{\Psi}({\bf r})= {\sum}^4_{i=1} \phi_i({\bf r})\,\hat{a}_i
\;
\eeq
obeying the standard bosonic commutators $[\hat{a}_{k},\hat{a}^{\dagger}_{i}]=\delta_{ki}$.
Owing to the form of the trapping potential, single-particle wave functions
$\phi_{k}({\bf r})$ exhibit a factorized form
\beq
\label{singleparticle}
\phi_{k}({\bf r})=g(z)\,w_k({\vec r}-{\vec r}_k)
\eeq
in which $g(z)$ represents the ground-state wave function of harmonic potential $(m \omega_{z}^2/2) z^2$,
and the planar wave function $w_k({\vec r}-{\vec r}_k)$ (${\vec r}_k$ is the center of the $k$th well) describes the localization at the $k$th well.

Note that we are assuming that the planar part of the potential $V_{t}$ is strong enough compared to other energies (the interaction energies in particular) such that the on-site wave functions $w_k({\vec r})$ ($k=1,...,4$) are fixed, being independent on the number of bosons in each well. We shall work under the hypothesis that the four minima of the $x-y$ potential are well separated. In such a way, the on-site wave function $w_k({\vec r})$ may be described by a single function $w_k=w ({\bf r} - {\bf r}_k)$, where ${\bf r}_k$ is the center of the $k$th well. The condition $w \ll \ell$ ($w$ is the width of each Gaussian in the $x-y$ potential of Eq. (\ref{quadruple})) entails that bosons are strongly localized in the proximity of sites ${\vec r}_k$ in the $x-y$ plane.
The functions $w_k({\vec r}-{\vec r}_k)$ and $w_i({\vec r}-{\vec r}_i)$ are orthogonal for $i \ne k$ so that one easily proves the orthonormality condition $\int d^{3}{\bf r}\phi_{k}^{*}({\bf r})\phi_{l}({\bf r})=\delta_{kl}$.

Potential $V({\bf r}-{\bf r'}) =$ $ g\, \delta^3 ({\bf r}-{\bf r'})+V_{dd}({\bf r}-{\bf r'})$,
describing boson-boson interactions, is the sum of a short-range (sr) $g$-dependent contact
potential (with $g=4\pi\hbar^2 a_s/m$ and $a_s$ the interatomic s-wave scattering length)
and of a long-range dipole-dipole (dd) potential
\beqa
\label{interaction}
V_{dd}({\bf r}-{\bf r'})=\gamma \frac{1-3 \cos^2 \theta}{|{\bf r}-{\bf r'}|^3}\,.
\eeqa
The coupling of dipoles through the relevant magnetic moment $\mu$ (electric moment $d$)
is embodied in $\gamma=\mu_0\mu^2/4\pi$ ($\gamma=d^2/4\pi \varepsilon_0$) in which
$\mu_0$ ($\varepsilon_0$) is the vacuum magnetic susceptibility (vacuum dielectric constant).
The relative position of the particles is given by the vector ${\bf r}-{\bf r'}$.
For external (electric or magnetic) fields large enough the boson dipoles are aligned along
the same direction, so that $\theta$ is the angle between the vector ${\bf r}-{\bf r'}$ and
the dipole orientation.

\subsection{4-well dipolar-boson model}

By assuming symmetric wells, the resulting dipolar-boson model is
described by the 4-site extended Bose-Hubbard (EBH) Hamiltonian
\beqa
\hat{H} = \hat{H}_I -J \, {\sum}^4_{i=1} \big(\hat{a}^{\dagger}_i \hat{a}_{i+1}+H.C.\big)
\label{4mode}
\;,\eeqa
where $J$ is the hopping amplitude, $\hat{a}_{i+4}= \hat{a}_{i}$ due to the ring geometry,
and the interaction Hamiltonian
\beqa
\hat{H}_I = {\sum}^4_{i=1} \left [ \frac{U_0}{2}  \hat{n}_i (\hat{n}_i-1)
+  U_1 \hat{n}_{i} \hat{n}_{i+1} \right ] \, ,
\label{HI}
\eeqa
in addition to the standard boson-boson interaction $U_0$, includes the
$U_1$-dependent term describing dipolar interactions. In $\hat{H}$,
the local number operator $\hat{n}_i = \hat{a}^{\dagger}_i \hat{a}_i$ counts
the number of particles in the $i$th well of the ring.

The three macroscopic parameters $J$, $U_0$ and $U_1$ are defined as follows.
The hopping amplitude is given by
\beq
\label{hopping}
J=-\int d^3{\bf r}\, \phi^{*}_{k}({\bf r})
\bigg[-\frac{\hbar^2}{2m}\nabla^2+V_{t}({\bf r})\bigg]\phi_{l}({\bf r})
\;,\eeq
where $k$ and $l$ ($k\neq l$) are two nearest-neighbor sites, while
the on-site interaction $U_0$ combines the contributions of
short-range and dipole-dipole interactions \cite{lahaye2}
\beqa
\label{onsite}
&&U_0= g \int d^3{\bf r}\, |\phi_k({\bf r})|^4\nonumber\\
&+&\frac{\gamma}{2} \int d^3{\bf r}\int d^3{\bf r'}\,
|\phi_k({\bf r})|^2V_{dd}({\bf r}-{\bf r'})\,|\phi_k({\bf r'})|^2
\; .
\eeqa
We write the nearest-neighbor interaction $U_1$ amplitude in the form
\beq
\label{densitydensity}
U_1= \frac{\gamma}{2} \int d^3{\bf r}  \int d^3{\bf r'}\,
|\phi_{k}({\bf r})|^2\,V_{dd}({\bf r}-{\bf r'})|\,|\phi_{l}({\bf r'})|^2
\; ,
\eeq
since the main contribution is due to the dipolar potential $V_{dd}$ \cite{lahaye2}.

The comparison with the 3-site extended BH model, describing dipolar bosons trapped
in three wells, shows how the presence of more than three wells induces significant
changes in the interaction processes.
In the triple-well case, the relevant Hamiltonian reads
$$
\hat{H}_3 = -J\big[\hat{a}^{\dagger}_1\hat{a}_2
 +\hat{a}^{\dagger}_2\hat{a}_1+\hat{a}^{\dagger}_2\hat{a}_3
 +\hat{a}^{\dagger}_3\hat{a}_2+\hat{a}^{\dagger}_1\hat{a}_3+
 \hat{a}^{\dagger}_3\hat{a}_1\big]\nonumber\\
$$
\vspace{-0.7cm}
\beqa
+\frac{U_0}{2} {\sum}^3_{i=1} \hat{n}_i (\hat{n}_i-1)
+U_1 \big[\hat{n}_1 \hat{n}_2+\hat{n}_2 \hat{n}_3+\hat{n}_1\hat{n}_3\big] \; .
\label{threemode}
\eeqa
Thanks to the equality
$$
\hat{N}^2={\sum}_{i=1}^{3}\hat{n}_{i}^{2}
+ 2(\hat{n}_1\hat{n}_2+\hat{n}_2\hat{n}_3+\hat{n}_1\hat{n}_3)\, ,
$$
where $\hat{N}=\hat{n}_1+\hat{n}_2+\hat{n}_3$ is such that $[\hat{H}_3,\hat{N}]=0$,
the Hamiltonian $\hat{H}_3$ reduces to the simpler 3-site BH model (the so-called BH trimer)
$$
\hat{H}_3 = -J\big[\hat{a}^{\dagger}_1\hat{a}_2
+\hat{a}^{\dagger}_2\hat{a}_1+\hat{a}^{\dagger}_2\hat{a}_3
+\hat{a}^{\dagger}_3\hat{a}_2+\hat{a}^{\dagger}_1\hat{a}_3+
\hat{a}^{\dagger}_3\hat{a}_1\big]\nonumber\\
$$
\beqa
\label{effectiveh}
+ \frac{U}{2}{\sum}^3_{i=1} \hat{n}_i (\hat{n}_i-1)
+\, \frac{U_1}{2} \hat{N}(\hat{N}-1)\; .
\eeqa
In the latter formula, $U \equiv U_0-U_1$ shows that the nearest-neighbor (dipolar) interactions
have been absorbed by the effective on-site interaction $U$ while $U_1$ only appears in the constant
term $U_1\hat{N}(\hat{N}-1)/2$. Based on this result, the ground-state structure of $H_3$
has been thoroughly investigated in Ref. \cite{dmps} by exploiting the well-known properties
of the BH-trimer ground state both in the attractive ($U<0$) and repulsive ($U>0$) interaction
regime.

The application of the same scheme to the 4-well dipolar model, where
$$
\hat{N}^2={\sum}_{i=1}^{4} \hat{n}_{i}^{2}
+ {\sum}_{i}{\sum}_{k\ne i} \hat{n}_i \hat{n}_k \, ,
$$
shows that, in addition to
$2(\hat{n}_1\hat{n}_2+\hat{n}_2\hat{n}_3+ \hat{n}_3\hat{n}_4+\hat{n}_4\hat{n}_1)$,
the nonlocal term depending on $\hat{n}_i \hat{n}_k$ now includes the contribution
$2( \hat{n}_1\hat{n}_3 +\hat{n}_2\hat{n}_4)$
involving the coupling of non adjacent opertors $\hat{n}_i$.
This leads to recast model (\ref{4mode}) into the form
\beqa
\label{ham4}
\hat{H} &=& C(\hat{N}) - J {\sum}_{i=1}^4 \big(\hat{a}^{\dagger}_i \hat{a}_{i+1}+H.C.\big)\nonumber\\
&+&\frac{U}{2}  {\sum}^4_{i=1} \hat{n}_i (\hat{n}_i-1)
-U_1 (\hat{n}_{1} \hat{n}_{3}+\hat{n}_{2} \hat{n}_{4} )
\;,\eeqa
where $U=U_0-U_1$ and $\displaystyle{C(\hat{N})= \frac{U_1}{2} \hat{N}^2 -\frac{U_1}{2} \hat{N}}$,
characterized by the extra term $\hat{n}_{1} \hat{n}_{3}+\hat{n}_{2} \hat{n}_{4}$
coupling next nearest-neighbor sites in the 4-site lattice. Then the new form (\ref{ham4})
of Hamiltonian (\ref{4mode}) is that of the 4-site BH model
where the effective on-site interaction parameter is once more $U = U_0-U_1$ but, unlike the case of
dipolar bosons in three-well potential, a new interaction term modify the spectral
properties of the model with respect to the case of the BH Hamiltonian.
\medskip

%
\section{The 4-well dipolar-boson model within the continuous variable picture}

The CVP 
is obtained by observing that physical quantities depending on the local boson populations
$n_i$ can be equivalently described in terms of densities $x_i = n_i /N$. For $N$ large enough,
the latter can be seen as continuous variables. This assumption leads to reformulate
in terms of densities $x_i$ both Fock states and, accordingly, the action of bosonic operators
on such states.
After setting $|n_1, n_2, ..., n_L \rangle \equiv |x_1, x_2, ..., x_L \rangle $ and observing
that creation (destruction) processes $n_i \to n_i +1$ ($n_i \to n_i -1$) entail that
$$
|x_1, ..., x_i ,... ,x_L \rangle \,\,\, \to \,\,\, |x_1, ..., x_i \pm \epsilon ,... ,x_L \rangle
\, , \quad \epsilon = 1/N\, ,
$$
one determines the effect of the action of Hamiltonian $\hat{H}$ on a generic quantum state
$|\Psi \rangle= {\sum}_n^* \Psi( {\vec n}  ) | {\vec n } \rangle$
where $|{\vec n } \rangle$ $= | n_1, n_2, \, ...n_i \, ...\rangle$.
The corresponding calculations describing the essence of this approach are discussed in {\ref{cvp}}.
%
Within the new formalism
the eigenvalue problem $\hat{H} |E \rangle = E |E \rangle $ for the BH Hamiltonian
$$
\hat{H} = \frac{U_0}{2} {\sum}^L_{i=1}  \hat{n}_i (\hat{n}_i-1)
-J{\sum}_{rs} A_{rs} \hat{a}^{\dagger}_r \hat{a}_{s}\, ,
$$
takes the CVP form \cite{vittorio2}
\beq
\Bigl [ -D + V \, \Bigr ]\, \psi_E ({\vec x})
 = \, {\bar E}  \, \psi_E  ({\vec x}) \, , \qquad {\bar E}= \frac{E}{N^2 |U|}\, ,
\label{cvp1}
\eeq
including the generalized Laplacian
$$
D = \, \tau  \sum_{rs}
\frac{\epsilon^2 }{2} A_{r s} \, \Bigl ( \partial_r -\partial_s \Bigr )\, {\sqrt { x_r \, x_s }}
\Bigl ( \partial_r -\partial_s \Bigr )\, ,
$$
with $\tau = J/(N |U|)$, and the effective potential
$$
V=
\frac{\sigma }{2} \sum^L_{r=1}   x_r (x_r  - \epsilon)
- 2\tau  \sum^4_{r=1} {\sqrt { x_r \, x_{r+1} }}\, ,
$$
where $\displaystyle{\sigma=U/|U|}$ assumes the value $\sigma=+1$ ($\sigma= -1$) in the
presence of an effective on-site repulsive (attractive) interaction $U>0$ ($U<0$).
 
The solutions $\psi_E ({\vec x})$ to this problem (and the relevant eigenvalues $E$) are
found by observing that it can be reduced to a multidimensional harmonic-oscillator problem
in the proximity of the extremal points of $V$. The essential information concerning the
ground-state configuration is thus obtained by imposing the stationarity condition of $V$. 
In the following, we use this condition to determine the ground state of model (\ref{ham4})
and its dependence on the model parameters.
%

\subsection{Bosonic-population equations charaterizing the ground state}

The application of the CVP to the 4-well dipolar-boson model (\ref{ham4})
yields the new eigenvalue equation
\beq
{\cal H}\, \psi_E ({\vec x}) = \, {\bar E}  \,\psi_E ({\vec x})
\, ,\qquad {\bar E}= \frac{E}{N^2 |U|}
\label{cvp2}
\eeq
where the effective Hamiltonian $\cal H$ contains the generalized Laplacian $D$
defined on the squared ring
$$
D = \, \tau \epsilon^2  \sum^4_{r=1} \, \Bigl ( \partial_{r+1} -\partial_r \Bigr )\, {\sqrt { x_r \, x_{r+1} }}
\Bigl ( \partial_{r+1} -\partial_r \Bigr )\, ,
$$
and the potential
$$
V=
\frac{\sigma }{2} \sum^4_{r=1}
x_r (x_r  - \epsilon)
- u_1 (x_{1} x_{3}+ x_{2} x_{4} )
- 2\tau  \sum^4_{r=1} {\sqrt { x_r \, x_{r+1} }}\, ,
$$
in which $u_1= U_1/|U|$ and $\sum_i x_i = 1$ owing to the conservation of the total boson number $N$.
At this point, we observe that due to ring geometry $x_{i+4}=x_i$ and for $N$ large enough the
parameter $\epsilon$ is sufficiently small. We can thus write a more useful version of $V$ which
underlines its symmetric character under exchanges of boson populations, that is

$$
V=  \frac{\sigma }{2} (x_1^2+x_2^2+x_3^2+x_4^2)
- u_1 (x_{1} x_{3}+ x_{2} x_{4} )
$$
\beq
- 2\tau
( \sqrt { x_2 } + \sqrt {x_4 } ) (\sqrt { x_3} + \sqrt {x_1 })\, ,
\label{pot}
\eeq
In general, two main regimes (the repulsive and attractive ones)
can be identified by considering the interplay between parameters $U_0$ and $U_1$
occurring in the effective potential $V$ or, equivalently, in Hamiltonian $\hat{H}$.
For a given $U_0$, one has
\beq
0 < U_1 < U_0 \quad \to \quad \sigma=+1 \, ,\,\,\, 0 < u_1 < \infty \, ,
\label{rep}
\eeq
%
%
%
\beq
U_0 < U_1 \quad \to \quad \sigma=-1 \, ,\,\,\, \infty > u_1 > 1\, .
\label{att}
\eeq
The derivation of the equations for variables $x_i$ discussed below shows how the
first case actually splits into two independent, significantly different, regimes.

To derive the equations for the $x_i$'s one must consider the constraint $1 = x_1+x_2+x_3+x_4$,
implying that one of the coordinates $x_i$ can be seen as a dependent variable. By assuming, for
example, $x_4 = 1-(x_1+x_2+x_3)$, the equations $\partial V /\partial x_i =0$ with
$i =1,2,3$ gives
$$
\sigma ( x_1 - x_4) -u_1 (x_3-x_2)
- \tau \frac{\sqrt x_4 + \sqrt x_2}{\sqrt x_1} + \tau \frac{\sqrt x_1 + \sqrt x_3}{\sqrt x_4} =0
$$
$$
( \sigma +u_1)( x_2 - x_4)
- \tau \frac{\sqrt x_1 + \sqrt x_3}{\sqrt x_2} + \tau \frac{\sqrt x_1 + \sqrt x_3}{\sqrt x_4} =0
$$
$$
\sigma ( x_3 - x_4) -u_1 (x_1-x_2)  -\tau \frac{\sqrt x_2+ \sqrt x_4}{\sqrt x_3}
+ \tau \frac{\sqrt x_1 + \sqrt x_3}{\sqrt x_4} =0
$$
determining the configurations $(x_1, x_2,x_3, x_4)$ for which
the stationarity condition of $V$ is realized. Since densities $x_i$ describe the
bosonic populations (BP), we will refer to such equations as the {\it BP equations}.
In general, such equations can be shown \cite{vittorio2} to describe
different weakly-excited states in addition to the ground state.
The solutions of the BP equations can be found analytically when one consideres
the special class of solutions for which $x_1 =x_3$. This reflects the exchange symmetry
$x_1  \leftrightarrow x_3$ characterizing such equations.
Then, by setting $x_1 =x_3$, one finds
\beq
\label{eqr1}
u_1 (x_1-x_2) - \sigma ( x_1 - x_4)
= \tau
\frac{2x_1-x_4-\sqrt x_2 \sqrt x_4}{\sqrt x_4 \sqrt x_1} \, ,
\eeq
\beq
\label{eqr2}
(\sigma +u_1) ( x_2 - x_4)
- 2\tau \frac{\sqrt x_1 }{\sqrt x_2} + 2\tau \frac{\sqrt x_1 }{\sqrt x_4} =0 ,
\eeq
showing how two of the three BP equations (the first and the third) are reduced to a unique equation.
The resulting system still contains the ground state. One easily checks that, with $\sigma = \pm 1$ and
for any value of $\tau$ and $u_1$, the uniform solution $x_2=x_4=x_1=x_3$ satisfies the
previous equations and reproduces the same ground state of the 4-site BH model in the
absence of dipolar interaction.

\subsection{Solutions of the BP equations}

A large amount of information can be extracted from the reduced BP equations. By rewriting
equation (\ref{eqr2}) in the form
\beq
\label{eqr22}
\Delta_{24} \left [  (\sigma +u_1) ( \sqrt x_2 + \sqrt x_4)
+ \frac{2\tau  \sqrt x_1 }{\sqrt x_2\sqrt x_4} \right ] =0\, ,
\eeq
with $\Delta_{24} = \sqrt x_2 - \sqrt x_4$,
one easily identifies the solution $x_2 =x_4$ entailing that the equation (\ref{eqr1}) becomes
\beq
\label{eqr12}
( x_1 - x_2) \left [ (\sigma -u_1)  + \frac{2\tau}{\sqrt x_1 \sqrt x_2} \right ] =0\, .
\eeq
Such equations show that three different regimes characterize the low-energy scenario
relevant to $V$ and, more in general, to the 4-well dipolar model.
For $U_1 < U_0$ (corresponding to $\sigma = +1$) one has two cases
$$
0< U_1 < U_0/2 \quad (\leftrightarrow \, u_1 <1)\,,
$$
$$
U_0/2 <U_1 < U_0 \quad (\leftrightarrow \, u_1 > 1)\, ,
$$
in addition to the case
$$
U_0 < U_1 < \infty \quad (\leftrightarrow \, u_1 > 1)\, ,
$$
where $\sigma = -1$. Note that, in the proximity of $U_0$ (namely, in the limit
$U_1 \to (U_0)^\pm$), parameter $u_1$ can assume arbitrarily large values.

Apart from $x_4= x_2$ and $x_2= x_1$,
equations (\ref{eqr22}) and (\ref{eqr12}) do not provide further solutions for $\sigma =+1$ and $u_1 <1$
in that the factors contained in the squared brackets is always positive.
Summarizing, the uniform solution $x_1= x_3=x_2=x_4$ is the unique solution
for $0 < U_1 < U_0/2$. This is the {\it ultraweak dipolar-interaction} regime.

A more structured solution is obtained from equation (\ref{eqr12})
in the interval $U_0/2 < U_1 < U_0$ where, in addition to $\sigma = +1$,
the inequality $u_1 >1$ holds. From equation (\ref{eqr12}) one gets
$$
x_1 x_2  = \frac{4\tau^2}{(u_1-1)^2}
$$
which, combined with the constraint $1 = 2(x_1+x_2)$, gives
\beq
x_1= \frac{1}{4} \left [ 1\pm \sqrt {1- f } \right ]\, ,\,\,\,
x_2= \frac{1}{4} \left [ 1\mp \sqrt {1- f } \right ]\, ,
\label{sol1a}
\eeq
with
\beq
f = \frac{64\tau^2}{(u_1-1)^2} = \frac{64 J^2}{N^2 (2U_1-U_0)^2} \, ,
\label{sol1b}
\eeq
in the interval $0 \le \tau \le |u_1-1|/8$. Then, by observing that $x_3=x_1$ and $x_4=x_2$,
the configuration of the system appears to be completely determined.
The range of $J$ where these solutions are defined depends on $U_0$ and $U_1$.
In view of definitions $u_1= U_1/|U_0-U_1|$ and $\tau = J/(N|U|)$),
after rewriting the latter inequality for $\tau$ in the form
\beq
J/N < |U_1-|U_0-U_1||/8,
\label{ineq1a}
\eeq
one finds that
for $U_1 \to U_0/2$ the range of $J/N$ tends to zero while, for $U_1 \to (U_0)^-$,
the range is $J/N < U_1/8$.
This case represents the {\it weak dipolar-interaction} regime.

A third case is found for $U_1 > U_0$ entailing $\sigma = -1$. Equations (\ref{eqr22})
and (\ref{eqr12}) take the form
\beq
\label{eqr23}
\Delta_{24} \left [  (u_1-1) ( \sqrt x_2 + \sqrt x_4)
+ \frac{2\tau  \sqrt x_1 }{\sqrt x_2\sqrt x_4} \right ] =0\, ,
\eeq
respectively.
\beq
\label{eqr13}
( x_1 - x_2) \left [ \frac{2\tau}{\sqrt x_1 \sqrt x_2} -(1+u_1)\, \right ] =0\, .
\eeq
Since $u_1 >1$, the first equation is solved once more by $x_2= x_4$ while the second one
is solved either by
$$
x_1= x_2 \quad ({\rm uniform \, solution})
$$
or by setting
$$
\frac{2\tau}{1+u_1} = \sqrt x_1 \sqrt x_2 \, .
$$
One immediately gets the relevant solutions given by
\beq
x_1= \frac{1}{4} \left [ 1\pm \sqrt {1- g} \right ]\, ,
\,\,\,
x_2= \frac{1}{4} \left [ 1\mp \sqrt {1- g} \right ]\, ,
\label{sol2a}
\eeq
with
\beq
g = \frac{64\tau^2}{(u_1+1)^2} = \frac{64J^2}{N^2 (2U_1-U_0)^2} \, ,
\label{sol2b}
\eeq
where the range of $\tau$ is determined by $0 \le \tau \le (u_1+1)/8$. Note that
$g\equiv f$ given by equation (\ref{sol1b}).
The inequality defining the upper bound can be rewritten in the more explicit form
$J\le N (2U_1-U_0)/8$ showing that $J \le NU_0/8$ for
$U_1 \to U_0^+$, and $T \le NU_1/8$ for a generic, arbitrarily large $U_1 > U_0$.
This case represents the {\it strong dipolar-interaction} regime.

\subsection{The low-energy scenario}

In the CVP form (\ref{cvp2}) of the dipolar-boson model
the energy of the system in the proximity of minimum-energy configurations
is described by potential (\ref{pot}).
The comparison of the energies corresponding
to the BP configurations analyzed in the previous section reveals the
change of structure of the minimum when the parameter $U_1$ is varied with respect
to $U_0$. Numerical calculations confirm that both the uniform solution
and solution (\ref{sol1a}) represent minimum-energy configurations
in the corresponding regimes.

The energy of the uniform solution $x_i = 1/4$ with $i= 1,2,3,4$ obtained from
(\ref{pot}) is easily found to be
$$
V_0 = \frac{\sigma -u_1}{8} - 2\tau \, , \quad \tau = \frac{J}{N|U|}\, .
$$
This represents the ground-state energy for $U_1 < U_0/2$, namely, when $u_1 <1$
and $\sigma = +1$ (ultraweak dipolar interaction).

In the subsequent interval $U_0/2 < U_1 < U_0$, where $\sigma = +1$ but $u_1 >1$,
the new solutions
(\ref{sol1a}), (\ref{sol1b}) have been found in addition to the uniform solution.
Since $x_1= x_3$ and $x_2= x_4$
$$
V =  (1- u_1) (x_1^2+x_2^2)- 8\tau\, \sqrt { x_2 } \sqrt {x_1 }
$$
which by using (\ref{sol1a}) and (\ref{sol1b}) gives
$$
V_0' = - \frac{|1- u_1|}{4} - \frac{ 8\tau^2 }{|1- u_1|}\, .
$$
For any value in $U_0/2 < U_1 < U_0$ one has
$$
V_0' = - \frac{|1- u_1|}{4} - \frac{ 8\tau^2 }{|1- u_1|} \, <\, V_0 = -\frac{|u_1-1|}{8} - 2\tau\, .
$$
By observing that inequality $\tau \le |1- u_1|/8$ defines the range of $\tau$,
the two energies are found to coincide for $\tau =|1- u_1|/8$, consistently with the fact
that, in this case, the solution described by (\ref{sol1a}) and (\ref{sol1b}) reduce to
the uniform solution.
For $U_1 \to (U_0/2)^+$ one has $u_1 \to 1^+$ which, owing to $\tau \le |u_1 -1|/8$,
implies that $\tau \to 0$. As expected, in this limit one obtains that $V_0 =0 = V_0'$.
Going to the opposite extreme $U_1 \to U_0^-$ one easily checks that
$$
V_0' = - \frac{U_0}{|U|}  \left ( \frac{1}{4} + \frac{ 8 J^2 }{N^2 U_0^2} \right )
\, <\,
V_0 = -\frac{U_0}{|U|}  \left (\frac{1}{8} + \frac{2J}{NU_0} \right )\, ,
$$
where inequality $\tau \le |1- u_1|/8$ is now substituted by $J \le N U_0/8$.
The condition $V_0' = V_0$ is reached for $J=NU_0/8$.
Note that the diverging factor $1/|U|$ in the previous expressions is in fact irrelevant
since the effective energies are defined by $N^2|U| \times V_0$
(see the energy eigenvalue in (\ref{cvp2})).

The transition from the regime $0 \le U_1 < U_0/2$ to the one with $U_0/2 < U_1 < U_0$ thus entails
the change of the ground state structure, which from the uniform-boson distribution
relevant to $x_i=1/4$ transforms into a non uniform distribution with two separated peaks
such that either $x_1=x_3 <x_2=x_4$ or $x_2=x_4 <x_1=x_3$.

For $U_1 \ge U_0$ (strong dipolar-iteraction regime) one has $\sigma = -1$ and $1 < u_1 < \infty$
for any $U_1$.  As a consequence, potential (\ref{pot}) takes the form
$$
V =  -(1+ u_1) (x_1^2+x_2^2)- 8\tau\, \sqrt { x_2 } \sqrt {x_1 }
$$
giving the minimum-energy formula
$$
V_0'' = - \frac{1+u_1}{4} - \frac{ 8\tau^2 }{1+u_1}\, ,
$$
when the $x_i$ corresponding to solution (\ref{sol2a}) are substituted.
Since
$$
1+ u_1 =
... =  \frac{2U_1 -U_0}{|U|}\, , \,\,\,
|1- u_1| =
... =  \frac{2U_1 -U_0}{|U|}\, ,
$$
in the interval $U_0/2 \le U_1\le U_0$ and $U_1\ge U_0$, respectively,
then $1+ u_1$ and $|1- u_1|$ describe the same function of $U_1$.
Hence, $V_0''$ simply represents the continuation of $V_0'$ in the upper interval
$U_1\ge U_0$ with $V_0''=V_0'$ for $U_1 \to U_0$. Once more, one easily checks that
$$
V_0'' = - \frac{1+ u_1}{4} - \frac{ 8\tau^2 }{1 +u_1} \, <\, V_0 = -\frac{u_1+1}{8} - 2\tau\, ,
$$
for essentially any value of $u_1$ where $V_0$ is the uniform-solution energy corresponding to the
choice $\sigma = -1$. Note that solution (\ref{sol2a}) is defined provided $\tau \le |1+u_1|/8$
is satisfied (the value $\tau \equiv (1+u_1)/8$
is the unique case for which $V_0'' =V_0$). Not surprisingly, this inequality reproduces
the more explicit one $J/N \le ({2U_1 -U_0})/(8{|U|})$ already found for solution (\ref{sol1a}).
For $U_1 \to (U_0)^+$ one has $J \le N U_0/8$ while for $U_1 >> U_0$ one has $J \le N U_1/4$.

The fact that $V_0'' =V_0$ for $J/N = ({2U_1 -U_0})/(8{|U|})$ suggests that when the
previous inequality is violated the minimum energy becomes that described by the uniform
solution. This circumstance is confirmed by the fact that both solution (\ref{sol1a})
and solution (\ref{sol2a}) reproduce the uniform solution whenever $\tau$ tends to its
extreme permitted value.

Concluding, this analysis shows that, rather counterintuitively, the crucial change
in the ground-state structure takes place when $U_1$ crosses $U_0/2$ (transition from
the ultraweak-interaction to the weak-interaction regime) while the change of the effective
dipolar interaction $U= U_1-U_0$ from repulsive ($U >0\, \to \sigma=+1$) to attractive
($U <0\, \to \sigma=-1$) is completely irrelevant.
The emerging ground state significantly differs from that of the dipolar-boson model
in a triple well. In the latter case the transition from the weakly-attractive regime,
characterized by the uniform solution (full boson delocalization), to the
strongly-attractive regime ($U_1 >> U_0$) shows that the ground state becomes a symmetric
superposition of three macroscopic states each one describing the almost complete
localization of bosons in one of the three wells (Schr\"odinger-cat state)
$$
|E_0 \rangle \simeq \frac{1}{\sqrt 2} \Bigl ( |N, 0 , 0 \rangle +
|0, N , 0 \rangle +|0, 0 , N \rangle \Bigr ) \, .
$$
Such a state manifestly reflects the equivalence of the 3-well dipolar-boson system
with the attractive BH trimer.
The 4-well dipolar model instead features a strongly-attractive regime where
bosons are macroscopically localized in non adjacent wells (for example, $x_1 =x_3 \simeq 1/2$
and $x_2 =x_4 \simeq 0$) and the ground state will be a symmetric superposition
$$
|E_0 \rangle \simeq \frac{1}{\sqrt 2} \Bigl ( |N/2, 0 , N/2, 0 \rangle +
|0, N/2, 0,N/2 \rangle  \Bigr )
$$
the second Fock state corresponding to the equivalent configuration $x_1 =x_3 \simeq 0$
and $x_2 =x_4 \simeq 1/2$). Even for $U_1 >> U_0$, states involving the full localization
of bosons in one of the four wells ($|N, 0 , 0, 0 \rangle$, $|0, N , 0, 0 \rangle$ , ... )
are in no way involved in the ground state. The latter in turn reflects the crucial role
played by the extra term $U_1 (n_1 n_3+n_2n_4)$ in the interaction Hamiltonian $\hat{H}_I$ in $\hat{H}$.
%

\section{Dipolar-boson ground state}

The ground state of the Hamiltonian (\ref{ham4}) can be written in the form of superposition of different Fock states which,
due to the conservation of total boson number $N$ reads
\beq
\label{superposition}
|\Psi\rangle=
\sum_{n_1=0}^{N}\,\sum_{n_2=0}^{N-n_1}\,\sum_{n_3=0}^{N-n_1-n_2}\,c_{n_1,n_2,n_3}\,|n_{1},n_{2},n_{3}\rangle
\;,
\eeq
where we have omitted the occupation number of the fourth well $n_4= N-(n_{1}+n_2+n_3)$.

\begin{figure}[h]
\centerline{\includegraphics[width=5cm,clip]{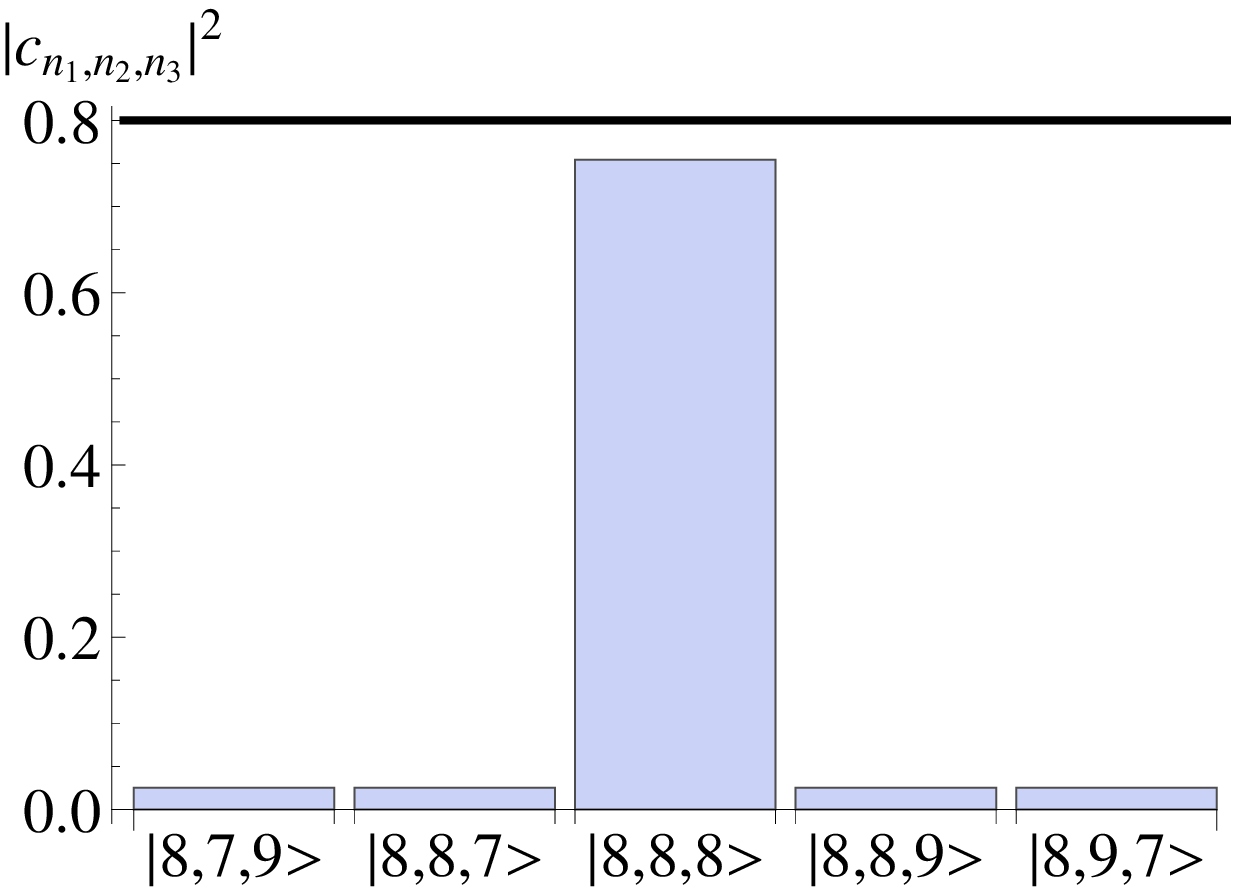}
\includegraphics[width=5cm,clip]{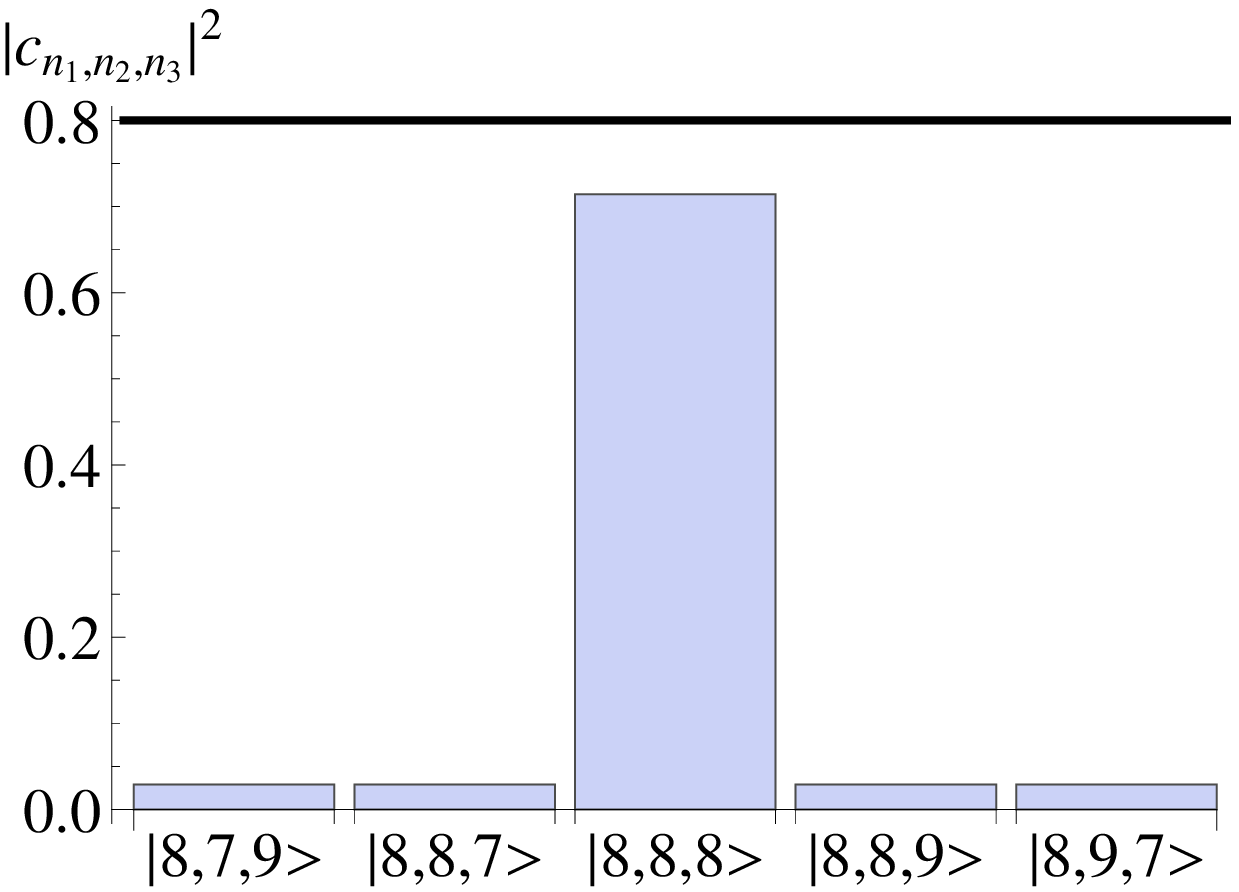}
\includegraphics[width=5cm,clip]{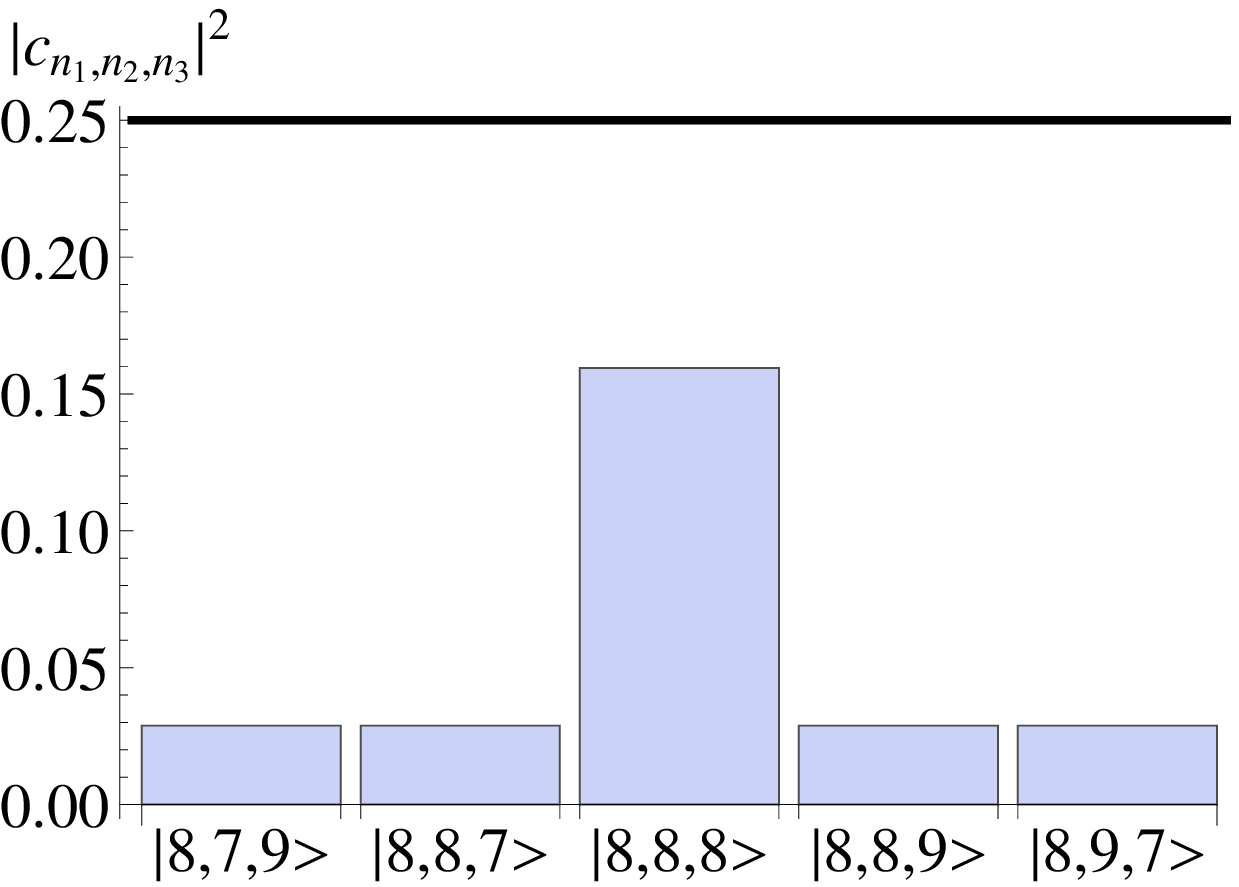}}
\centerline{
\includegraphics[width=5cm,clip]{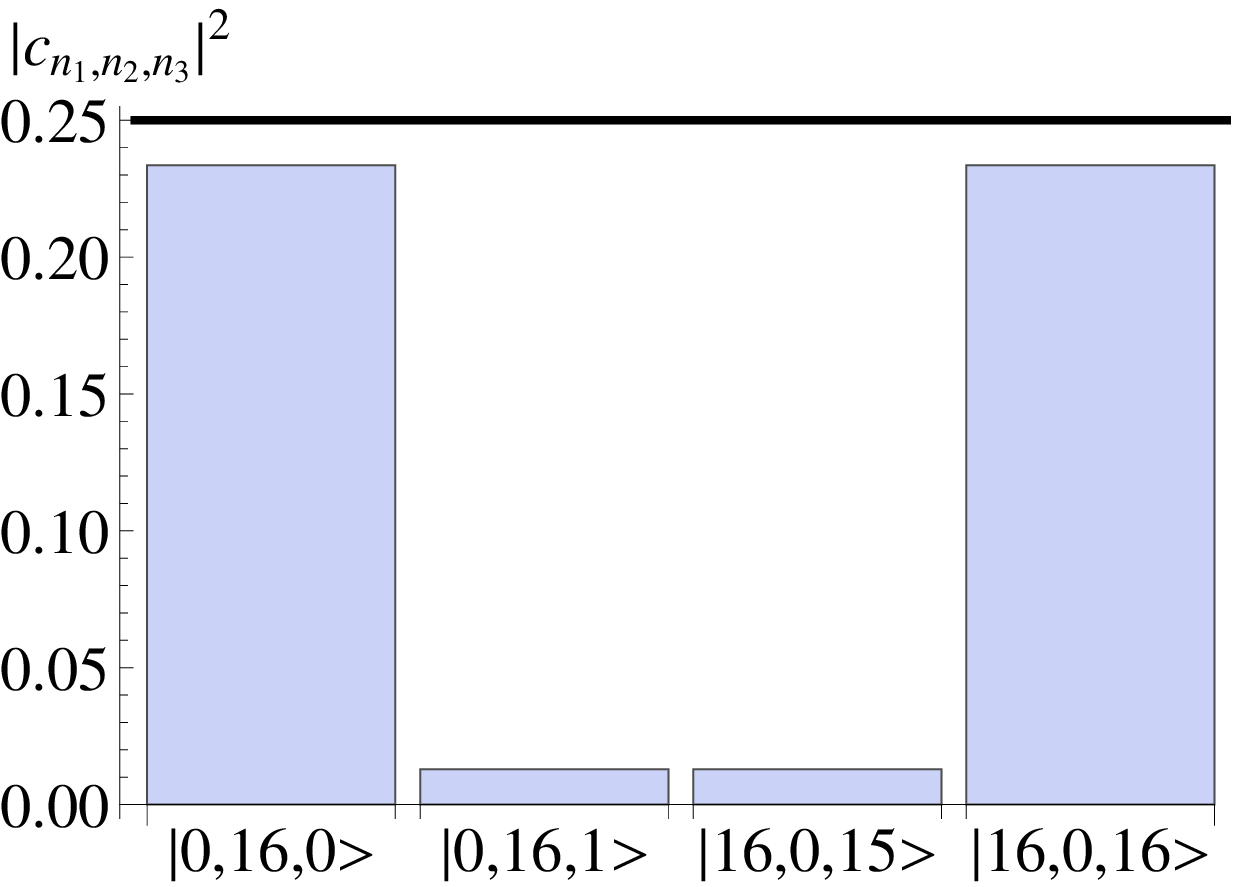}
\includegraphics[width=5cm,clip]{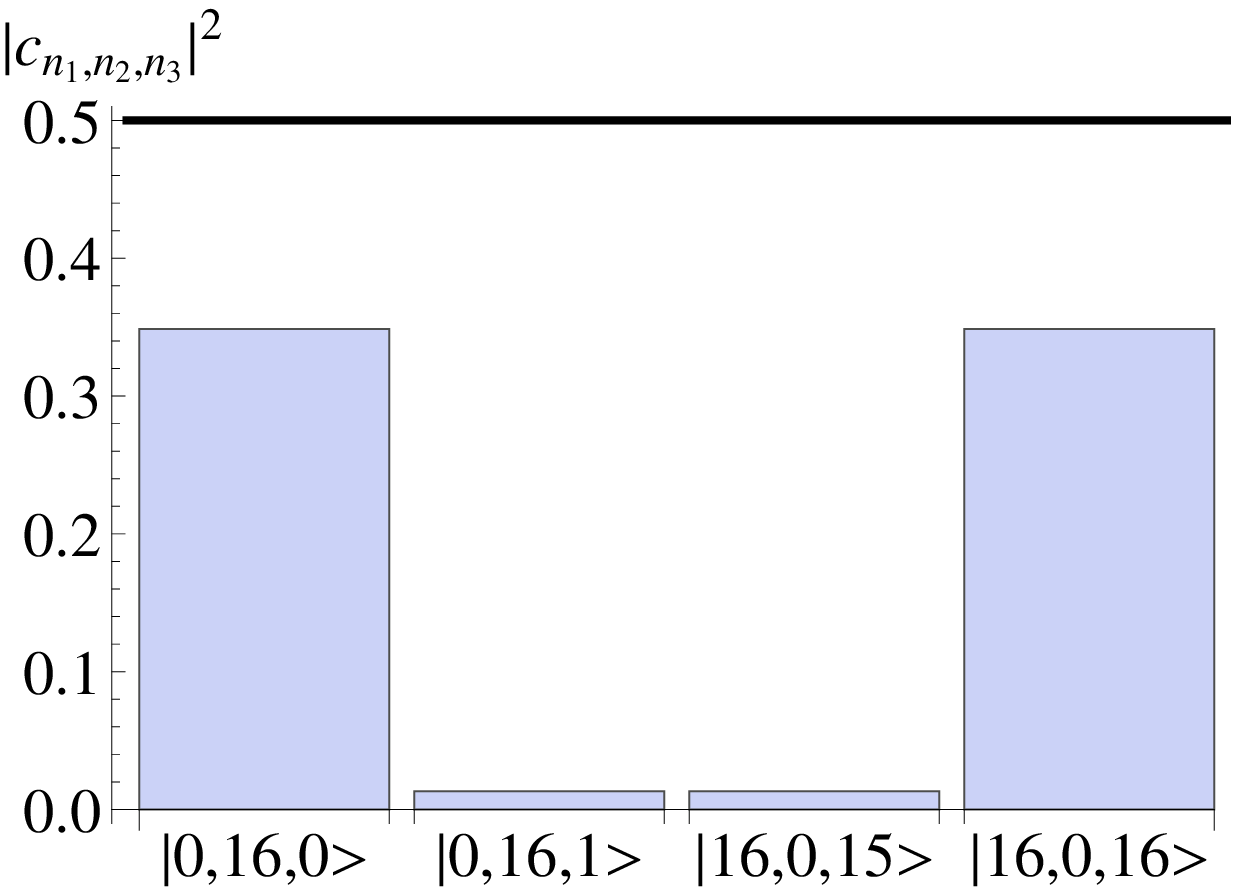}
\includegraphics[width=5cm,clip]{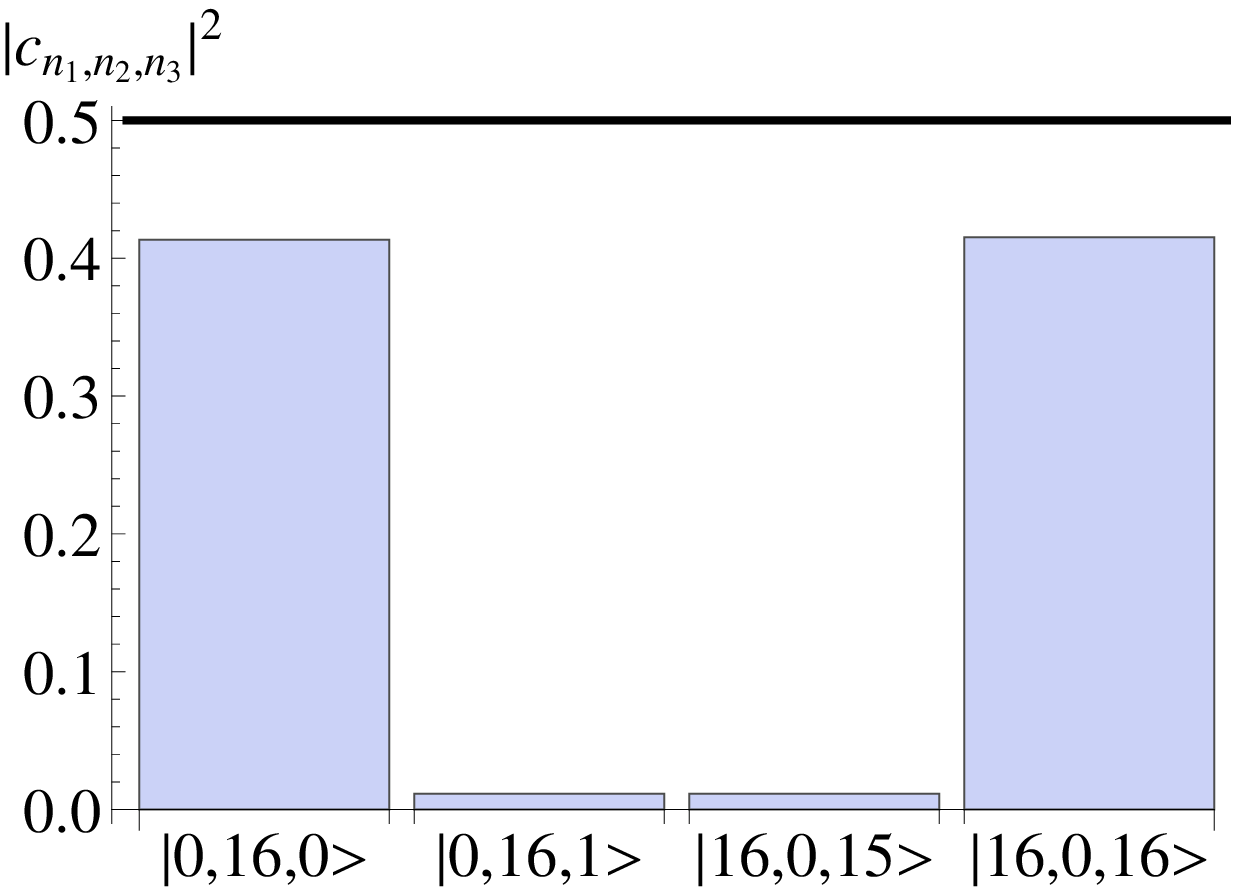}
}
\caption{(Color online). Horizontal axis: kets $|n_1,n_2,n_3\rangle$ [the fourth well occupation being $n_4=N-(n_1+n_2+n_3)$]. Vertical axis: the squared modulus of ground-state coefficients $|c_{n_1,n_2,n_3}|^2$. Here: $U_0=1$, $J=0.02$ and $N=32$. Top panels: $0<U_1<U_0/2$. First (from the left): $U_1=0.00001$. Second: $U_1=0.1$. Third: $U_1=0.499$. Bottom panels: $U_0/2<U_1<U_0$. First (from the left): $U_1=0.50251$. Second: $U_1=0.5030$. Third: $U_1=0.5038$. In the bottom panels, parameters lie in the filled triangle plotted in Fig.~\ref{fig2}.}
\label{fig1}
\end{figure}




\begin{figure}
\epsfig{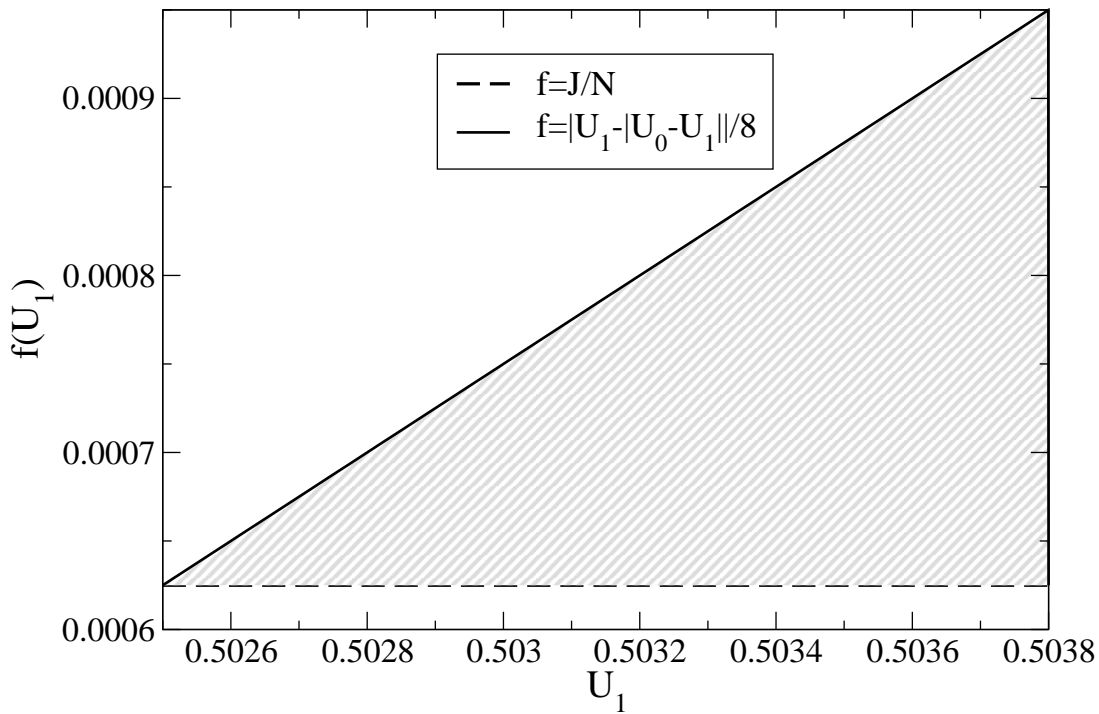}
\caption{The shaded area is the region where the solutions $x_1$ and $x_2$ 
(given by Eq. (\ref{sol1a})) are defined: the inequality (\ref{ineq1a}) 
is satisfied. Dashed line: left-hand-side of Eq. (\ref{ineq1a}). 
Solid line: right-hand-side of Eq. (\ref{ineq1a}). $U_0=1$, $J=0.02$, $N=32$.}
\label{fig2}
\end{figure}

\begin{figure}[h]
\centerline{\includegraphics[width=5cm,clip]{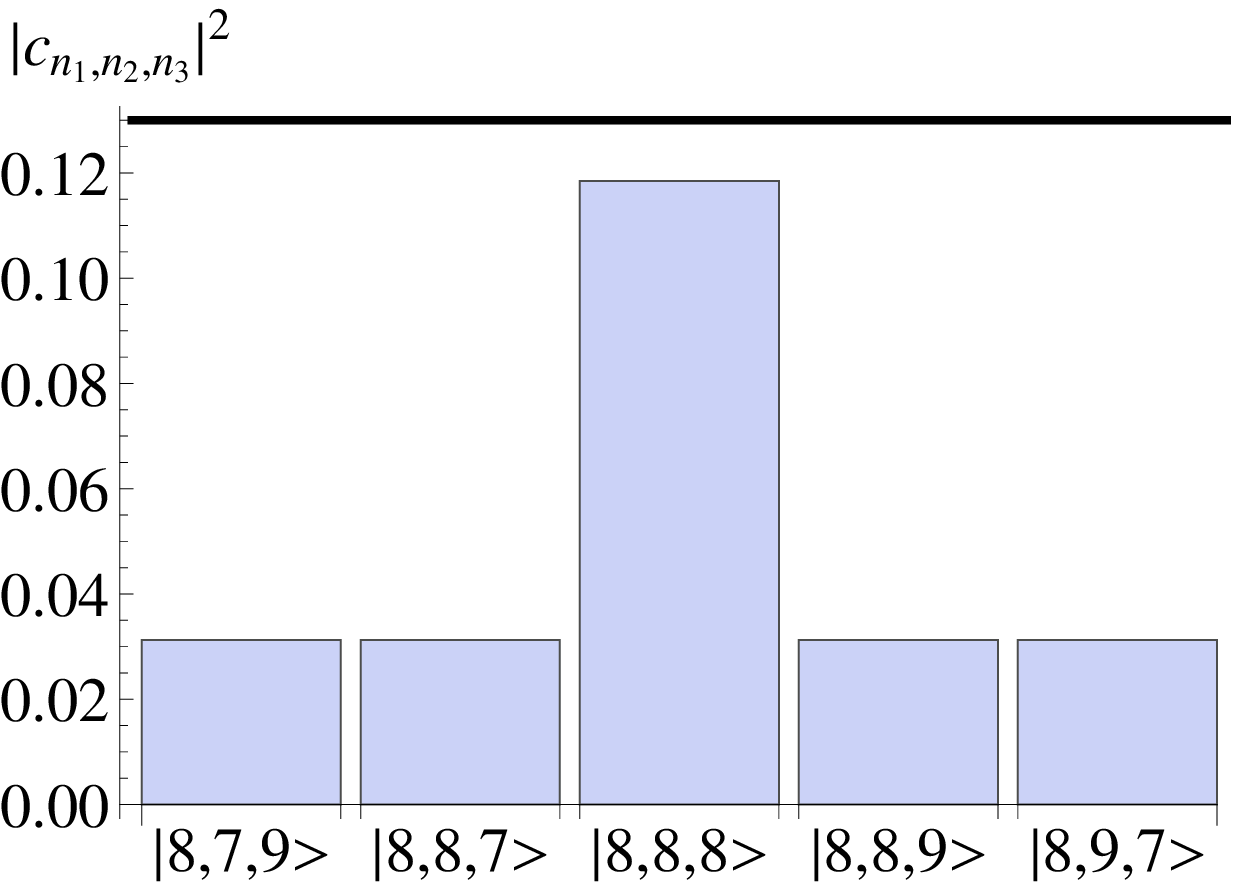}
\includegraphics[width=5cm,clip]{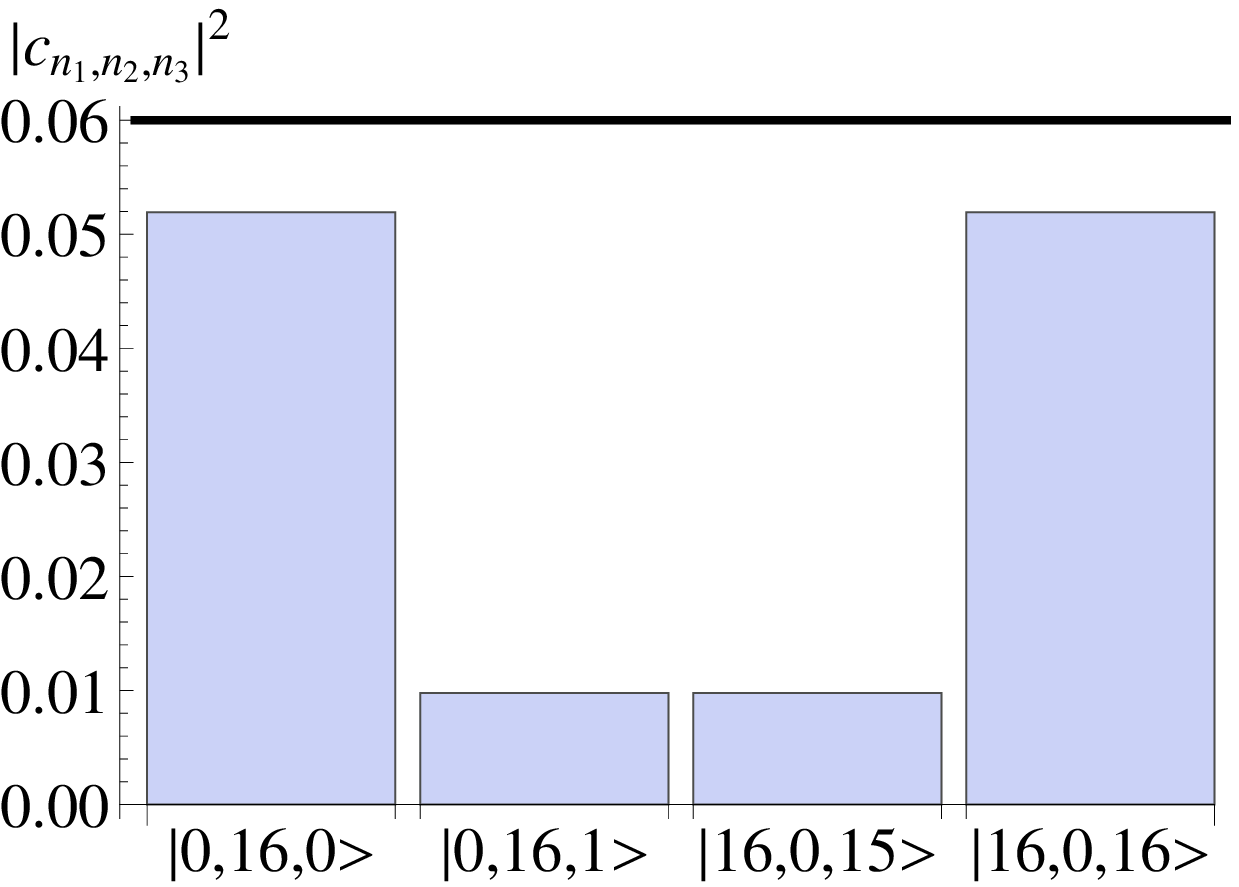}
}
\centerline{\includegraphics[width=5cm,clip]{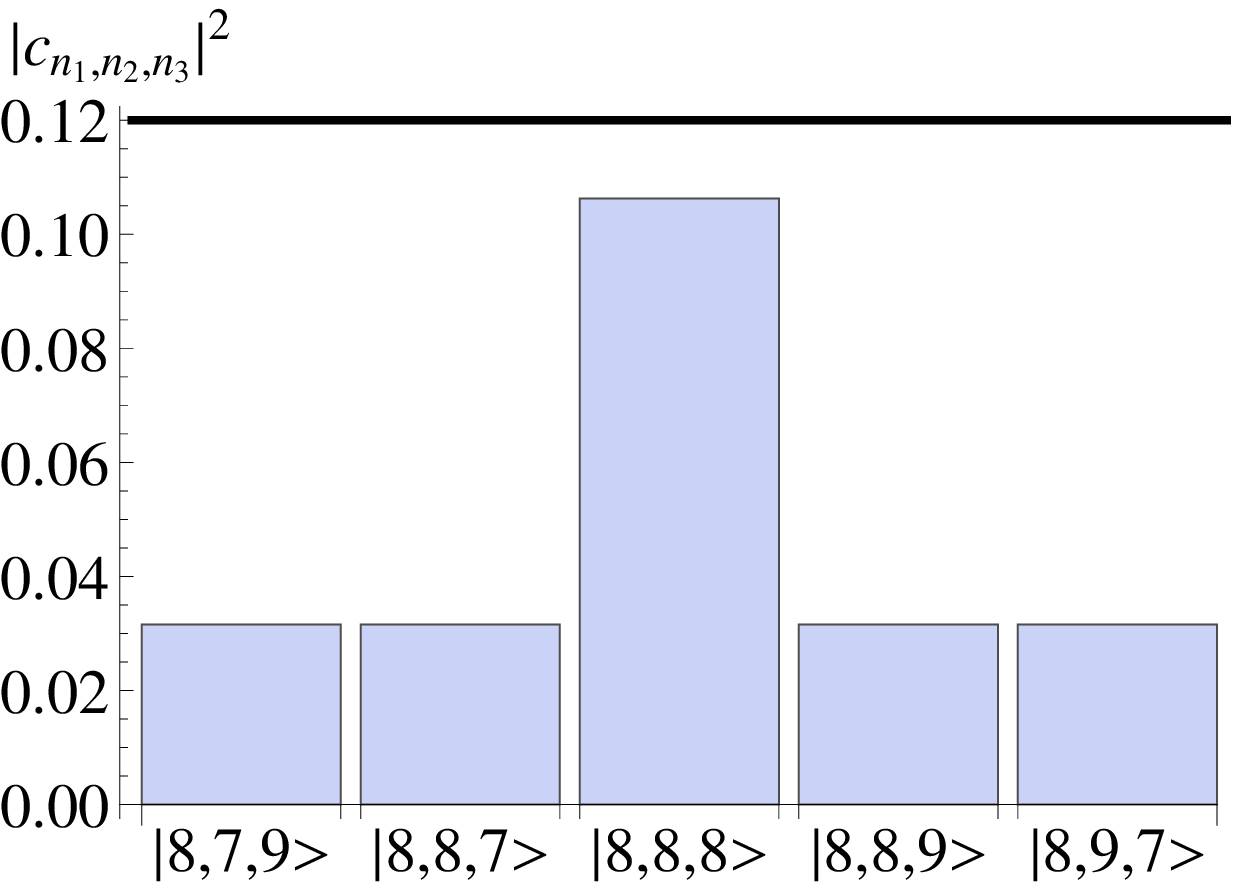}
\includegraphics[width=5cm,clip]{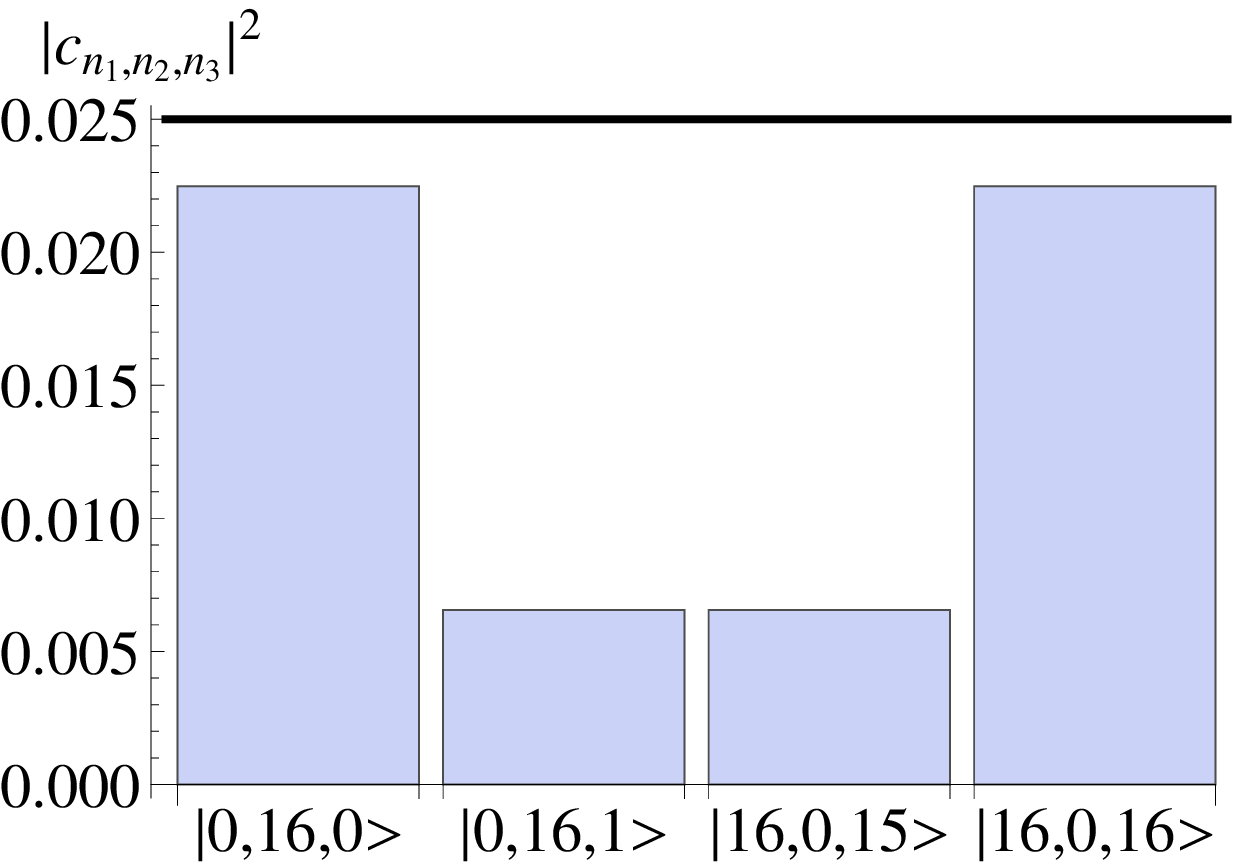}
}
\centerline{\includegraphics[width=5cm,clip]{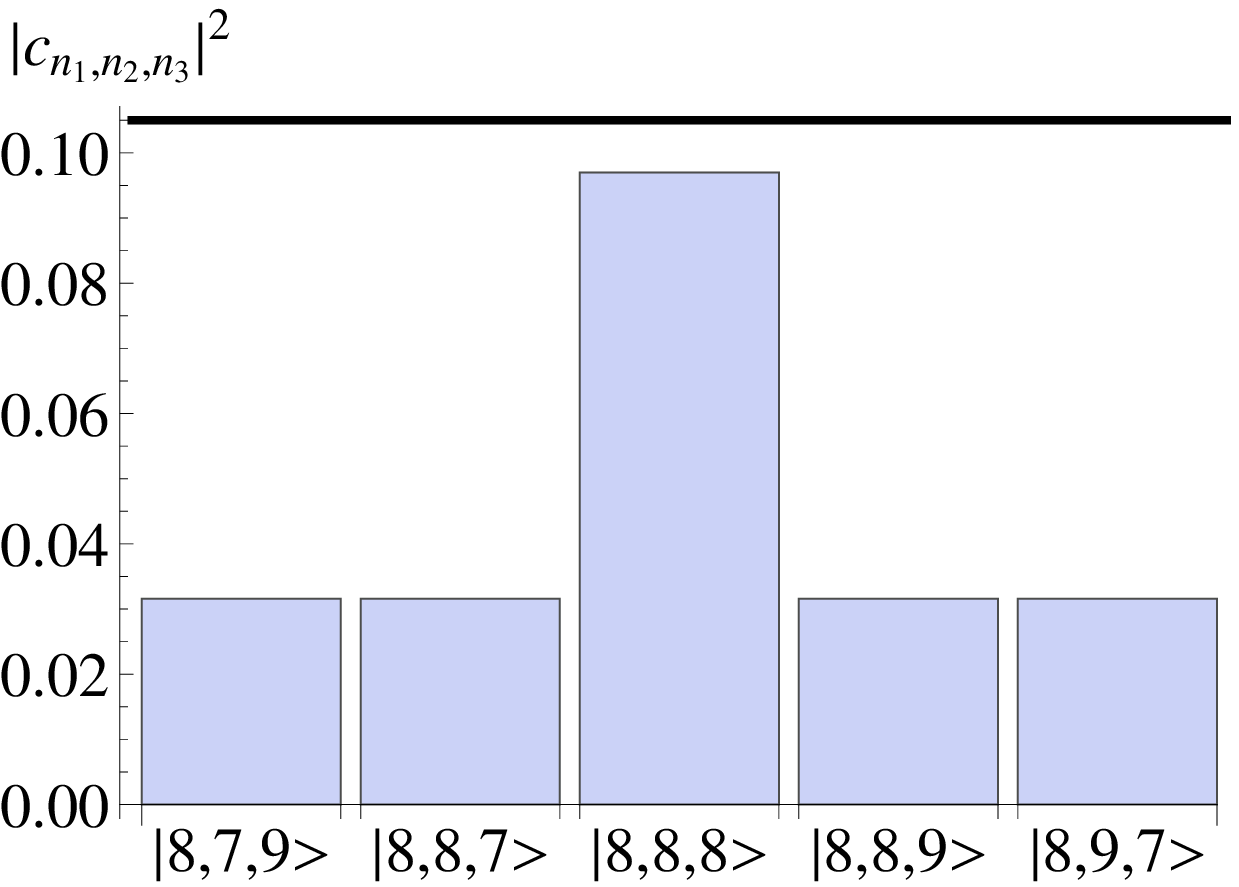}
\includegraphics[width=5cm,clip]{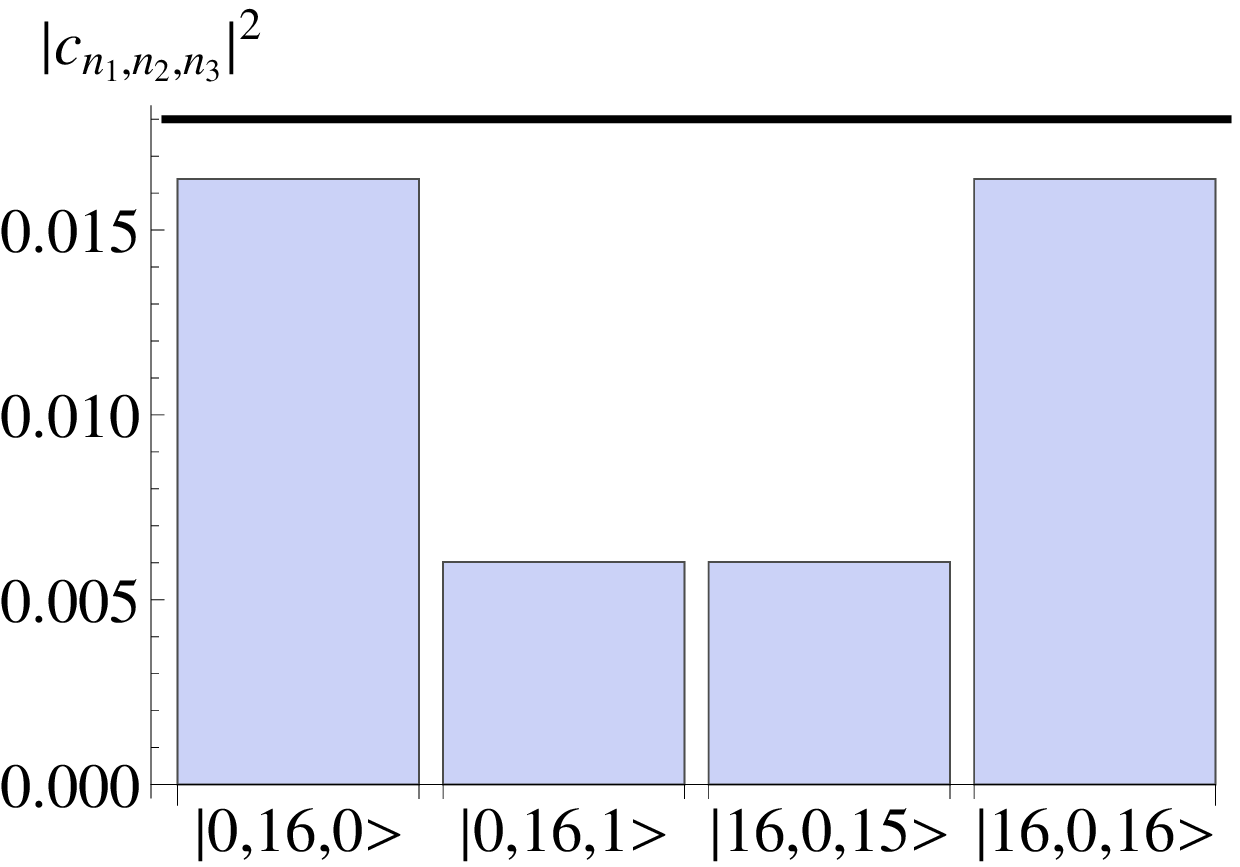}
}
\caption{(Color online). Horizontal axis: kets $|n_1,n_2,n_3\rangle$ 
[the fourth well occupation being $n_4=N-(n_1+n_2+n_3)$]. Vertical axis: 
the squared modulus of ground-state coefficients $|c_{n_1,n_2,n_3}|^2$. 
Here: $U_0=1$ and $N=32$. Top panels ($J=0.03$): $U_1=0.499$ (left);
$U_1=0.50438$ (right). Middle panels ($J=0.035$): $U_1=0.499$ (left); 
$U_1=0.5038$ (right). Bottom panels ($J=0.04$): $U_1=0.499$ (left); 
$U_1=0.5051$ (right).}
\label{fig3}
\end{figure}


Different ground states $|\Psi\rangle$ are sustained by the Hamiltonian (\ref{ham4})
depending on the relative magnitude of the parameters $U_1$ and $U_0$. To show how this causes
changes in the ground state structure, we have studied the probabilities $|c_{n_1,n_2,n_3}|^2$
on varying of $U_1$ in suitable ranges of values (see the discussion below) by keeping fixed
both $J$ and $U_0$. The results of this analysis, performed for a total boson number $N=32$, are shown in Fig.~\ref{fig1} (where, for convenience of representation, we have reported only $|c_{n_1,n_2,n_3}|^2 \gtrsim 1\times 10^{-2}$). Moreover, in Figs. ~\ref{fig1} and ~\ref{fig2} we assume $U_0$ as the energy scale ($U_0=1$).
Let us give a look to plots of Fig.~\ref{fig1} starting from the top panels. Here $0<U_1<U_0/2$. This corresponds to the ultraweak dipolar-interaction regime. By inspecting these distributions, two things can be clearly observed.
As a first, $|c_{n_1,n_2,n_3}|^2$ attains its maximum value for $n_1=n_2=n_3=8$ that is equivalent,
in the CVP language, to $x_1=x_2=x_3=x_4$: the uniform solution. The second observation is that
increasing $U_1$ has the effect to produce a progressive depletion of state $|8,8,8\rangle$,
i.e. state $|N/4,N/4,N/4,N/4\rangle$ remains the maximally populated one,
thus confirming the predictions of the CVP approach.

The bottom panels of Fig.~\ref{fig1}, instead, represent what happens in the weak
dipolar-interaction regime characterized by $U_0/2<U_1<U_0$. We have chosen the values of $U_1$
in accordance with the analysis of the previous section so that to satisfy the inequality
(\ref{ineq1a}). We have studied, in other words, the ground state corresponding to the parameter
values in the (lower) shaded area of Fig.~\ref{fig2}. The panel corresponding to $U_1=0.50251$ reveals that a transition occurs
in the ground state: the bosons populate with the highest probability the states $|0,16,0\rangle$
and $|16,0,16\rangle$ (i.e. the states $|0,N/2,0,N/2\rangle$ and $|N/2,0,N/2,0\rangle$, respectively)
that, in the CVP fashion, correspond to $x_3=x_1$ and $x_4=x_2$.

The ground state of the Hamiltonian (\ref{ham4}) is therefore a symmetric superposition of the states $|0,N/2,0,N/2\rangle$ and
$|N/2,0,N/2,0\rangle$. This result corroborates the CVP studies that predict a change in the ground state structure from the uniform state to a macroscopic two-pulse state when $U_1$ crosses $U_0/2$. By further increasing $U_1$ the above superposition is still the ground state
and the $|c|^2$'s  pertaining to $|0,16,0\rangle$ and $|16,0,16\rangle$ becomes larger,
as it can be seen from the fifth and sixth panels corresponding to $U_1=0.5030$ and $U_1=0.5038$, respectively.

As a conclusive remark, we note that the CVP-predicted localization-delocalization transition 
is captured by numerics in wider terms.
In fact, we have found numerically the ground-state of the four-site BH Hamiltonian (\ref{ham4}) 
with $N=32$ in correspondence to $J=0.03$ (top panels of Fig.~\ref{fig3}) and 
$J=0.035$ (middle panels of Fig.~\ref{fig3}), and $J=0.04$ (bottom panels of Fig.~\ref{fig3}).  
From these plots, one can clearly see the change experienced by the ground-state 
when the boundary $U_0/2$ is crossed, like so expected from CVP. As for Fig.~\ref{fig1}, 
also in this case, the second panel of each $J$-fixed pair has been obtained by choosing 
the Hamiltonian parameters the inequality (\ref{ineq1a}) is satisfied.


\section{Conclusions}
We have considered a system of interacting dipolar bosons confined by a four-well potential
with a ring geometry. The microscopic dynamics of this system are ruled by a four-site
Bose-Hubbard (BH) model including interactions between bosons in next-nearest-neighbor wells.
We have studied the ground state of the 4-well realization of the BH Hamiltonian by varying the
amplitude $U_1$ of the interaction of next-nearest-neighbor wells.

We have attacked the problem from two sides, i.e. both analytically and numerically. From
the analytical point of view we have reformulated the dipolar-boson model within the framework
of the continuous variable picture (CVP). By exploiting this approach we have shown that the
ground state structure exhibits a dramatic change when the amplitude $U_1$ becomes larger than
a precise fractional value of the on-site interaction $U_0$. More precisely,
the condition $U_1 >U_0/2$ signs the delocalization-localization transition. In the delocalization
regime, the ground state is uniform (equally shared bosons among the four wells), while in the
localization one, the system is a macroscopic two-pulse state where the bosons are strongly
localized. These CVP results are corroborated by those obtained from the numerical diagonalization
of the four-site BH Hamiltonian. Indeed, within this approach it can be clearly observed that in
the delocalization regime the ground state is, practically, a Fock state with equally
occupied sites, whereas the localization regime corresponds to a symmetric superposition
of two (macroscopic) Fock states each one describing non adjacent wells populated by the half
of the bosons in the system. In contrast with dipolar bosons in double- and triple-well
potentials \cite{mazzdell,dmps}, even when $U_1 \gg U_0$, there are no ground state
that can be represented as a symmetric superposition of Fock states involving the
full localization of bosons in one of the four sites. This result, in addition to validate
the approach based on the CVP to the low-energy states of many-bosons systems, shows the
considerable influence of the number of wells on the ground-state structure
and prompts further study on dipolar bosons trapped in a ring lattice involving
$L>4$ potential wells.\\

%

This work has been supported by MIUR (PRIN Grant No. 2010LLKJBX).
GM acknowledges financial support from the University of Padova (Progetto di Ateneo Grant No. CPDA118083) and Cariparo Foundation (Eccellenza Grant 2011/2012).
GM thanks P. Buonsante, S.M. Giampaolo, and M. Galante for useful comments and suggestions.

\bigskip
\bigskip
\bigskip

\appendix
\section{Application of the CVP}
\label{cvp}

The action of the hopping-term operators on a generic quantum state
$|\Psi \rangle= {\sum}_n^* \Psi( {\vec n}  ) | {\vec n } \rangle$
where $|{\vec n } \rangle$ represents the Fock state $| n_1, n_2, \, ...n_i \, ...\rangle$, yields
$$
\hat{a}^{\dagger}_{s} a_r |\Psi \rangle
=
{\sum}_x^* \psi( {\vec x}  ) N {\sqrt { x_r }} {\sqrt { x_s +\epsilon }}\,
|..., x_r -\epsilon,..., x_s +\epsilon, ...\rangle
$$
$$
=
N {\sum}_x^* \psi( x_1, ..., x_r +\epsilon, ..., x_s -\epsilon , ...)
{\sqrt { x_r + \epsilon }} {\sqrt { x_s }}\, | {\vec x } \rangle\, ,
$$
where $\psi( {\vec x}  ) \equiv \psi( x_1, x_2 ..., x_L)$ has replaced $\Psi( {\vec n} )$.
In this scheme, the key approximation amounts to assume that
only the first and second-order contributions must be considered in the
Taylor expansion in $\epsilon$ of the function
$\psi(..., x_r +\epsilon, ..., x_s -\epsilon , ...) {\sqrt { x_r + \epsilon }} {\sqrt { x_s }}$
occurring in $\hat{a}^{\dagger}_{s} a_r |\Psi \rangle $.
This gives
$$
\hat{a}^{\dagger}_{s} a_r |\Psi \rangle
=
N {\sum}_x^*
\Bigl [  {\sqrt { x_r \, x_s }} \psi( x)  +
\, \epsilon {\sqrt { x_r \, x_s }}\, \left (\frac{\partial \psi}{\partial x_r}\,
- \frac{\partial \psi}{\partial x_s} \right )
$$
$$
+
{\sqrt { x_r \, x_s }} \frac{\psi({\vec x}) }{2x_r} \, \epsilon
+
\frac{\epsilon^2}{2} {\sqrt { x_r \, x_s }}
\Bigl ( \frac{\partial^2 \psi}{\partial x^2_r} + \frac{\partial^2 \psi}{\partial x^2_s}\,
- 2 \frac{\partial^2 \psi}{\partial x_r\partial x_s}
\Bigr )
$$
$$
+ {\sqrt { x_r \, x_s }} \left [
\, -\frac{\psi( x) }{8x^2_r}
+
\frac{1 }{2x_r} \left ( \frac{\partial \psi}{\partial x_r}\,
- \frac{\partial \psi}{\partial x_s} \right ) \right ] \, \epsilon^2
+ ...\Bigr ]\, |{\vec x}\rangle\, .
$$
Then, the action of the typical hopping term of BH Hamiltonians
$\sum_{s ,r} A_{sr} \hat{a}^{\dagger}_{s} a_r $ on a generic state $|\Psi \rangle $
can be shown to be represented by
$$
\sum_{s ,r} A_{sr} a^+_{s} a_r |\Psi \rangle
= \frac{ N }{2} \sum^*_{\vec n}  \Bigl \{
\sum_{s ,r}  A_{rs}  \psi({\vec x})
\times
$$
$$
\Bigl [ 2 \sqrt { x_r \, x_s }
+
\frac{\epsilon }{2} \frac{x_r + x_s}{ {\sqrt { x_r  x_s }} }
- \frac{\epsilon^2 }{8} \left ( \frac{\sqrt { x_r } }{\sqrt { x_s^3}}
+ \frac{\sqrt { x_s } }{\sqrt { x_r^3}} \right )  \Bigr ]
\,
$$
$$
+ \epsilon^2  A_{rs} (\partial_{x_r} -\partial_{x_s})
{\sqrt { x_r  x_s }} (\partial_{x_r} -\partial_{x_s}) \psi({\vec x}) \Bigr \}
\, |{\vec x}\rangle
\, .
$$
In such formulas $A_{rs}$ represents the adjacency matrix. The latter is, in general, zero except for
nearest-neighbor sites for which $A_{rs}= 1$.
To complete the description of this scheme, one must consider
the action of terms such as $\sum_i \hat{n}_i^k $ on $|\Psi\rangle$.
This is easily found to be
$$
\sum_i \hat{n}_i^k |\Psi\rangle = \, N^{-k}\sum_i x_i^k \psi({\vec x})  |{\vec x}\rangle
\, .
$$
\vfill \eject

\end{document}